\def\Box{\kern1pt\vbox{\hrule height 1.2pt\hbox{\vrule width 1.2pt\hskip 3pt 
\vbox{\vskip 6pt}\hskip 3pt\vrule width 0.6pt}\hrule height 0.6pt}\kern1pt} 
\def\gtwid{\mathrel{\raise.3ex\hbox{$>$\kern-.75em\lower1ex\hbox{$\sim$}}}} 
\def\ltwid{\mathrel{\raise.3ex\hbox{$<$\kern-.75em\lower1ex\hbox{$\sim$}}}} 
\def\be{\begin{equation}} 
\def\ee{\end{equation}} 
\def\beq{\begin{eqnarray}} 
\def\eeq{\end{eqnarray}} 
 
\documentstyle[epsfig,12pt]{article} 
 
\begin{document} 
\begin{titlepage} 

\begin{center} 
{\Large \bf On the stability of gravity in the presence of a non-minimally coupled scalar field}
\end{center} 
 
\vskip 0.5cm 
 
\begin{center} 
 L. Raul Abramo$^{\dagger}$ \\ 
 \footnotesize{Instituto de F\'{\i}sica, Universidade de S\~ao Paulo,\\ 
 CP 66318, 05315-970 S\~ao Paulo, Brazil, and\\ 
 
 Theoretische Physik der Universit\"at M\"unchen, \\ 
 Theresienstr. 37, D-80333 M\"unchen, Germany,} 
\end{center} 
\begin{center} 
 L\'eon Brenig$^\ddagger$ \\ 
 \footnotesize{RggR, Universit\'e Libre de Bruxelles,\\  
 CP 231, 1050 Bruxelles, Belgium, and \\ 
 
 Service de Physique Statistique, Universit\'e Libre de Bruxelles,\\  
 CP 231, 1050 Bruxelles, Belgium, 
} 
\end{center} 
\begin{center} 
 and Edgard Gunzig$^\star$\\ 
 \footnotesize{RggR, Universit\'e Libre de Bruxelles,\\  
  CP 231, 1050 Bruxelles, Belgium, and\\ 
 
  Instituts Internationaux de Chimie et de Physique Solvay,\\  
  CP 231, 1050 Bruxelles, Belgium} 
\end{center} 
 
\vskip 2.0cm  
 
\begin{center} 
ABSTRACT 
\end{center} 
\hspace*{.5cm}
We show that Einstein's gravity coupled to a non-minimally
coupled scalar field is stable even for high values of the
scalar field, when the sign of the Einstein-Hilbert action 
is reversed. We also discuss inflationary solutions and a 
possible new mechanism of reheating.

\begin{flushleft} 
\end{flushleft}

\vspace{2cm} 
\begin{flushleft} 
 $^{\dagger}$ e-mail: abramo@fma.if.usp.br \\ 
 $^{\ddagger}$ e-mail: lbrenig@ulb.ac.be \\ 
 $^{\star}$ e-mail: egunzig@ulb.ac.be 
\end{flushleft} 
\end{titlepage}

\noindent{\bf Introduction} 

\vskip 0.3cm 
 
\noindent Non-minimal couplings of a scalar field to spacetime curvature 
appear very frequently in theoretical physics, in 
applications ranging from alternative 
theories of gravity \cite{GRTests} to attempts to quantize gravity \cite{QG} 
to, more recently, scalar field models of ``dark energy'' \cite{NMC_Q,Farese}.
The implications of non-minimal coupling were also extensively
investigated in connection with inflation \cite{NMC_Inf}.

Recently, some of us studied a previously unexplored 
sector of Einstein's gravity with a non-minimally coupled (NMC) scalar, 
and found a class of new inflationary solutions \cite{Gunzigetal}. 
A distinguishing feature of this sector is that the coupling 
to curvature becomes so important that it leads to new features
such as a graceful dynamics from flat space to inflation 
and ending in a Friedman-Robertson-Walker spacetime, and ``superinflation'' 
(i.e. $dH/dt > 0$, where $H$ is the expansion rate). However, ominously, 
the model also brings a dynamical reversal of the sign of the 
gravitational action.

This last feature should alarm those familiar with General 
Relativity: the ``wrong'' sign for the Einstein-Hilbert action 
means that the excitations of the gravitational field
have negative-energy modes.
Therefore, this ``sign-reversal'' of the action could be
a hint that the theory is unstable in that 
sector \cite{Farese}, since the positive-energy 
scalar field would feed the negative-energy gravitons and  
vice-versa, leading to an explosive process.

Non-minimal coupling then seemed to be in a predicament: on the 
one hand it is an unavoidable interaction \cite{RevGunzig,Whyxi}
which produces a very attractive cosmology \cite{Gunzigetal}.
On the other hand, the theory appears to be unstable
just in the sector which contains the interesting new dynamics.

This Letter shows that in fact there is no predicament:
Einstein's theory of gravity with a scalar field 
is in fact stable on all sectors for $\xi \leq 0$ and $\xi \geq 1/6$.
The stability of the theory follows from the fact that, for this range
of $\xi$, both the pure gravitational {\it and} the pure scalar degrees 
of freedom of the theory simultaneously reverse the signs of their
actions. This opens the way to novel inflationary scenarios, and 
may even suggest a new mechanism for reheating the Universe after 
inflation. In a forthcoming paper \cite{ABG02} we show 
that the inflationary model gives phenomenologically sound predictions about 
the strength and scale-dependence of the spectrum of anisotropies 
of the cosmic microwave background radiation.

\vskip 0.7cm
{\noindent \bf Stability of Einstein's theory with a scalar field}
\vskip 0.3cm

\noindent The system is Einstein's General Relativity
with the addition of a scalar field:
\be
\label{L_CC} 
L \equiv \sqrt{-g} \left\{ - \frac12 \left( 1 - \xi \psi^2 \right) 
R +  \frac12 g^{\mu\nu} \psi_{, \mu} \psi_{, \nu} - V(\psi) \right\} \; , 
\ee 
where in our conventions $8 \pi G = 1$ and the metric has
timelike signature $(+,-,-,-)$. $V(\psi)$ is an arbitrary scalar
self-interaction potential. The scalar coupling to curvature
$\xi$ can in principle assume any value: $\xi=0$ corresponds to
no (``minimal'') coupling to curvature, while $\xi=1/6$ is the case of
``conformal'' coupling.

The ``conformal factor'' $F (\psi) \equiv 1 - \xi \psi^2$ in the
Lagrangean (\ref{L_CC})
multiplies the usual Einstein-Hilbert term $- \sqrt{-g} R/2$.
Because the scalar field multiplies the scalar curvature, it is clear that
the physical (helicity two) degrees of freedom of the gravitational 
field are intertwined with the (helicity zero) degrees of freedom 
of the scalar field.

The diagonalization of the system (\ref{L_CC}) can be achieved 
through a conformal transformation of the metric:
\be
\label{cg} 
\tilde{g}_{\mu\nu} = \Omega^2 g_{\mu\nu} \; .
\ee 
This change of variable induces the following transformation 
in the Ricci scalar (in four dimensions):
\be
\label{cR} 
\tilde{R} \equiv R \left[ \tilde{g} \right] =  
\Omega^{-2} \left[ R - 6 g^{\mu\nu} D_\mu D_\nu \left( \log{\Omega} \right) 
- 6 g^{\mu\nu} D_\mu \left( \log{\Omega} \right)  
D_\nu \left( \log{\Omega} \right) \right] \; , 
\ee 
where $D_\mu[g]$ are covariant derivatives.
Using this expression in Eq. (\ref{L_CC}) we obtain, after
neglecting total derivatives:
\be 
\label{L2} 
L_> = \sqrt{-\tilde{g}} \left\{ -
\frac{F(\psi)}{2\Omega^2} \tilde{R} + \frac12 \tilde{g}^{\mu\nu} 
\left( \frac{\psi_{,\mu} \psi_{,\nu}}{\Omega^2}  
+ 6 \frac{\Omega_{,\mu}}{\Omega} \frac{\Omega_{,\nu}}{\Omega} \right) 
- \frac{V(\psi)}{\Omega^4} \right\} \; .
\ee

We want to rewrite this Lagrangian in such a way that it 
resembles as much as possible the Einstein-Hilbert action 
plus a minimally coupled scalar field \cite{Bek}. Hence, let us take 
$\Omega^2 = F(\psi)$ assuming for the moment that $F(\psi) > 0$,
which leads to:
\be 
\label{L3} 
L_> = \sqrt{-\tilde{g}} \left\{ 
- \frac12 \tilde{R} + \frac12 \tilde{g}^{\mu\nu} \psi_{,\mu} \psi_{,\nu}
\frac{1+\xi(6\xi-1)\psi^2 }{F^2(\psi)} 
- \frac{V(\psi)}{F^2(\psi)} \right\} \; . 
\ee
If we now introduce the conformally transformed field and effective
potential
\be
\label{tpsi} 
d \tilde{\psi} \equiv
\frac{\sqrt{1+\xi(6\xi-1)\psi^2}}{F(\psi)} d \psi \quad , \quad
\tilde{V} \equiv \frac{V(\psi)}{F^2(\psi)} \; , 
\ee
and substitute these expressions into Eq. (\ref{L3}) we obtain: 
\be 
\label{Lt} 
L_> = \sqrt{-\tilde{g}} \left\{  
- \frac12 \tilde{R} +  
\frac12 \tilde{g}^{\mu\nu} \tilde\psi_{,\mu} \tilde\psi_{,\nu} 
- \tilde{V} \right\} \; . 
\ee 
Because the Lagrangian (\ref{Lt}) has the form of the Einstein-Hilbert 
action plus a scalar term, it is common to call the tilded variables 
the ``Einstein frame''. Conversely, the original action (\ref{L_CC}) is 
referred to as the action in the ``Jordan frame''. 
Notice the minus sign in front of the Einstein-Hilbert action both
in the Jordan as well as in the Einstein frames: it
guarantees, in our conventions, that in the ultraviolet 
(or geometrical optics) limit
gravitons behave as positive-energy free fields \cite{MTW}.

Of course, for the transformation (\ref{tpsi}) to make sense
the terms inside the square root should be positive, that is:
\be
\label{ineq}
1 + \xi (6\xi-1) \psi^2 \geq 0 \; .  
\ee
For $F (\psi) \geq 0$ it is easy to check that this condition is 
always satisfied.

However, $F$ can be negative {\it if} $\xi \geq 0$ {\it and}
the value of the scalar field is sufficiently high ($\psi^2 > 1/\xi$).
I this case, we would like to use a conformal transformation of the 
metric which preserves its original signature,
so we take $\Omega^2 = - F > 0$. With this choice we have 
the counterpart to Eq. (\ref{cR}): 
\be 
\label{cRt} 
\tilde{R} \equiv R \left[ \tilde{g} \right] =  
- \frac{1}{F} \left[ R - 6 g^{\mu\nu} D_\mu D_\nu  
\left( \log{\Omega} \right) 
- 6 g^{\mu\nu} D_\mu \left( \log{\Omega} \right)  
D_\nu \left( \log{\Omega} \right) \right] \; . 
\ee 
Neglecting total derivatives once again, we obtain an 
expression very similar to Eq. (\ref{L2}): 
\be 
\label{L2t} 
L_< = \sqrt{-\tilde{g}} \left\{ 
\frac12 \tilde{R} - \frac12 \tilde{g}^{\mu\nu} 
\left[ \frac{\psi_{,\mu} \psi_{,\nu}}{F}  
+ \frac32 \frac{F_{,\mu}}{F} \frac{F_{,\nu}}{F} \right] 
- \frac{V(\psi)}{F^2} \right\} \; .
\ee 
Again, the term between square brackets simplifies:
\be 
\label{L3t} 
L_< = \sqrt{-\tilde{g}} \left\{ 
\frac12 \tilde{R} - \frac12 \tilde{g}^{\mu\nu} \psi_{,\mu} \psi_{,\nu}
\frac{1+\xi(6\xi-1)\psi^2}{F^2} 
- \frac{V(\psi)}{F^2} \right\} \; . 
\ee
Notice that both the Einstein-Hilbert and the scalar kinetic terms
have a switched sign with respect to (\ref{L3}), as opposed to 
the scalar potential term which does not.
We therefore use the field redefinitions:
\be 
\label{tpsit} 
d \tilde{\psi} \equiv
\frac{\sqrt{1+\xi(6\xi-1)\psi^2}}{F(\psi)} d \psi \quad , \quad
\tilde{V} \equiv - \frac{V(\psi)}{F^2(\psi)} \; , 
\ee 
Substituting these expressions into Eq. (\ref{L3t}) we obtain: 
\be 
\label{Ltt} 
L_{<} = - \sqrt{-\tilde{g}} \left\{  
- \frac12 \tilde{R} + 
\frac12 \tilde{g}^{\mu\nu} \tilde\psi_{,\mu} \tilde\psi_{,\nu} 
- \tilde{V} \right\} \; . 
\ee 

The crucial fact to notice now is that the Lagrangian (\ref{Ltt})
has an overall minus sign that appeared as a consequence of the
change of variables with $F<0$. Therefore, the excitations of both the 
scalar {\it and} gravitational fields carry energy with the same sign, 
even though the sign happens to be negative in this case. But an overall
sign in front of the Lagrangian is irrelevant, so the theory is
in effect stable for $F<0$ \footnote{Of course, the sign of the 
gravity-scalar Lagrangian is only irrelevant if this theory exists by
itself: if other matter fields are included, the relative signs of 
{\it their} Lagrangians should be the same as that of the gravity-scalar
sector in order to avoid instabilities.}.

We should again take care that the field transformation
(\ref{tpsit}) is well-defined, and the condition is still given
by inequality (\ref{ineq}). For $\xi \leq 0$ and $\xi \geq 1/6$ that 
condition is always satisfied if $F \leq 0$. However, for $F<0$ and
$0<\xi<1/6$ the condition is violated when the scalar field lies 
outside the range $1 \leq \xi \psi^2 \leq 1/(1-6\xi)$.
In that case, the scalar kinetic term in the Lagrangian (\ref{L3t})
reverses its sign. What this means is that for $0<\xi<1/6$ there 
is an {\it unstable sector} of the theory corresponding to large 
values of the scalar field\footnote{Similar conditions were also reached
by \cite{Hosotani}.}.

The consequences of a kinetic term which can reverse its sign 
depending on the sector of the theory are quite severe:
this introduces negative-energy states which make the whole
theory highly unstable.
Even if the value of the field is in the stable
sector in a certain region of space initially, the existence of the 
unstable sector means that the scalar field will tunnel from the
stable sector into the unstable sector.
It is highly doubtful whether such a theory could exist
even for a moment.

This means that the cases of mininal coupling, $\xi=0$, 
and of conformal coupling, $\xi=1/6$, are threshold systems:
for the sake of the global stability of gravity with a scalar field,
either $\xi \leq 0$ or $\xi \geq 1/6$. The case of minimal coupling case
is stable simply because the excitations of the gravitational
and scalar fields are always positive. The conformal coupling
case is stable because the excitations of the 
gravitational and scalar fields both carry the same 
sign --- positive or negative.

There is also a trivial method by which the ``effective'' Lagrangian
(\ref{Ltt}) could be derived in the case of negative $F$: suppose 
that we let the metric change its signature 
if $F$ changes sign, so that the  
conformal transformation into the Einstein frame is $\Omega^2=F$ in
any case. This corresponds
to a change of the lorentzian signature --- from timelike
to spacelike or vice-versa, depending on the convention.
The Lagrangian 
would still be expressed exactly as in Eq. (\ref{Lt}). However, 
the fact that the signature of the metric is switched with respect 
to the original definition means that both $\tilde{R}$ and 
$\tilde{g}^{\mu\nu} \tilde\psi_{,\mu} \tilde\psi_{,\nu}$ have 
in fact switched signs with respect to the original (positive) Lagrangian.
If we wanted to restore the usual signature, we would have to let 
$\tilde{g}_{\mu\nu} \rightarrow - \tilde{g}_{\mu\nu}$, and the 
signature restoration would lead precisely to Eq. (\ref{Ltt}), with the 
definitions (\ref{tpsit}).

There are, however, two problems with this global description of
a NMC scalar field plus gravity.
First, that it ignores the rest of the world (we will discuss
this point at the conclusion.)
And second, that the scalar potential
remains unchanged while all other terms of the Lagrangian
change sign, so its r\^ole is reversed if the signature changes.

\vskip 0.7cm

{\noindent \bf Scalar self-interactions, inflation and superinflation}

\vskip 0.3cm 

\noindent We henceforth consider only minimal and
conformal couplings, since all other cases are qualitatively
equivalent to either one of those.
The case of minimal coupling is of course trivial: any potential 
with a lower bound gives a physically sensible
(i.e., stable) theory.

However, in the conformally coupled case, if $F<0$ ($\psi^2 > 6$) 
then by Eq. (\ref{tpsit}) the
Einstein frame potential becomes $\tilde{V} = - V/F^2$. Hence, 
a physical (Jordan frame) scalar potential which is bounded from 
below but unbounded from above for large values of the scalar field
could become, in the $F<0$ sector of 
the Einstein frame, bounded from above but unbounded from below.
We therefore come to the conclusion that in the case of
Einstein's gravity with a conformally coupled scalar field, the
large scalar field sector must have a Jordan-frame potential which is
bounded from {\it above}, and not necessarily bounded from below!

Take the simplest scalar potential:
\be
\label{pot}
V(\psi) = \frac12 m^2 \psi^2 - \frac{\lambda}{4} \psi^4 \; ,
\ee
where $\lambda>0$ in the conformally coupled case.
The Einstein-frame
scalar field is obtained by direct integration of Eq. 
(\ref{tpsi}):
\be
\label{psipsi}
\tilde{\psi} = \left\{ 
\begin{array}{lll} 
\sqrt{6} \tanh^{-1}{} \frac{\psi}{\sqrt{6}} \quad , 
\quad  \psi^2 < 6 \; , \\
\\
\sqrt{6} \tanh^{-1}{} \frac{\sqrt{6}}{\psi} \quad , 
\quad  \psi^2 > 6 \; .
\end{array} \right.
\ee
(Notice that $\tilde\psi$ is {\it not} monotonic in $\psi$.)
The potential in the Einstein frame is, by
Eqs. (\ref{tpsi}) and (\ref{tpsit}):
\be
\label{tv}
\tilde{V} = \left\{ 
\begin{array}{lll} 
3 m^2 \cosh{\frac{\tilde\psi}{\sqrt{6}}} \sinh{\frac{\tilde\psi}{\sqrt{6}}}
\left( 1 - \frac{3 \lambda}{m^2} \tanh{\frac{\tilde\psi}{\sqrt{6}}} \right)
\quad , \quad  \psi^2 < 6 \; , \\
\\
- 3 m^2 \cosh{\frac{\tilde\psi}{\sqrt{6}}} \sinh{\frac{\tilde\psi}{\sqrt{6}}}
\left( 1 - \frac{3 \lambda}{m^2} \coth{\frac{\tilde\psi}{\sqrt{6}}} \right)
\quad , \quad  \psi^2 > 6 \; .
\end{array} \right.
\ee
The potentials are plotted in Fig. 1, where the left and right panels 
correspond respectively to
the sectors $F>0$ ($\psi^2 < 6$) and $F<0$ ($\psi^2 > 6$).
For the range of parameters 
$1/6 < \lambda/m^2 < 1/3 $, the potential has a maximum in 
the $F<0$ sector (right panel, solid line.)
This corresponds to a de Sitter fixed point in the Einstein frame, 
which is itself the image of a de Sitter fixed point in the Jordan frame
in the $F<0$ sector first exhibited in \cite{Gunzigetal}. 
Notice that the Einstein-frame potential in the $F<0$ sector becomes
exponentially negative when $\tilde{\psi} \rightarrow \infty$ 
($\psi \rightarrow \sqrt{6}$). This is of course due to the fact that
$F(\psi)$, which appears in the denominator of $\tilde{V}$, vanishes
when $\psi = \sqrt{6}$.

\begin{figure} 
\centering\epsfig{file=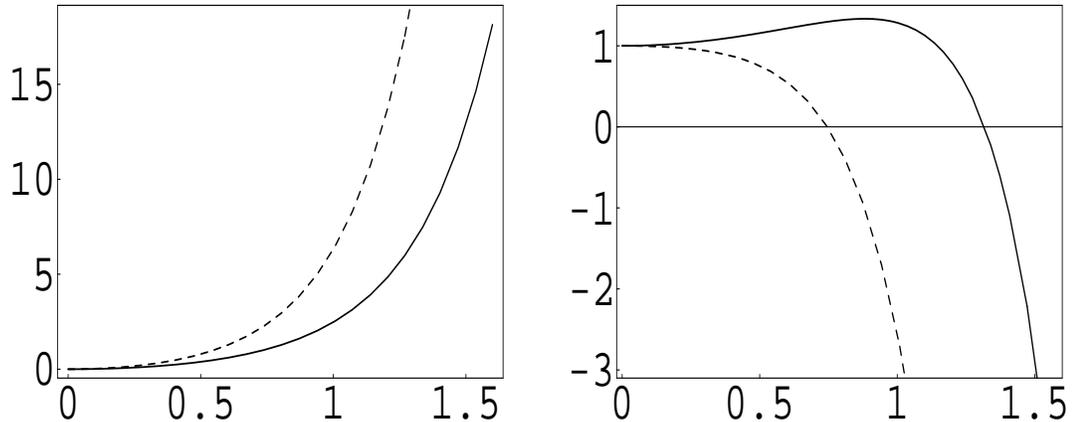,height=6.0cm,width=15cm}
\caption{\footnotesize  Einstein-frame scalar potential $\tilde{V}$
as a functon of $\tilde\psi$ in the $F>0$ (left) 
and $F<0$ (right) sectors. The solid and dashed lines correspond 
respectively to $\lambda/m^2 = 1/4$ and $\lambda/m^2 = 1/8$.}
\label{fig1}
\end{figure}

It is interesting to consider what would happen if the scalar runs
to the right of the fixed point at the maximum of the potential in
the right panel of Fig. 1. Under the approximation of an 
exponential potential, the solutions for a homogeneous scalar 
field in Einstein's gravity are those of power-law inflation \cite{PowerLaw},
and in this particular case what would happen is that the field
$\tilde{\psi} \rightarrow \infty$ (or, $\psi \rightarrow \sqrt{6}$) 
in a finite time. But then the sector changes from $F<0$ to $F>0$, and
the field finds itself in the potential of the left panel of Fig. 1.
The field then runs down to the bottom of the potential (again in a finite 
time), and relaxes there leading to a mild Friedman-Robertson-Walker 
expansion \cite{Gunzigetal}.

The problem with this particular scenario is that although
the homogeneous solutions in the Jordan frame are perfectly finite and 
well-behaved as $F \rightarrow 0$ \cite{Gunzigetal}, 
inhomogeneous fluctuations 
diverge in a few cases, as the system approaches that point 
\cite{Starob,ABG02}. 
The cause of these divergences is 
that the infinitely negative values of the ``effective'' 
potential $\tilde{V}$ are reached over a finite period of time.
Although the background quantities remain finite, because
the divergence factors out from the diagonal Einstein's 
equations, this is not {\it a priori} true for the perturbations,
which are determined by the full Einstein's equations 
(more precisely, it is the anisotropic stress which is causing
the divergence.)

Divergences can be avoided by taking a potential
$\tilde{V}$ which is well-behaved everywhere --- i.e., bounded from 
below at $|\tilde{\psi}| < \infty$.
But then the transition between sectors $F<0$ and $F>0$ seem unlikely.
In a forthcoming paper we will discuss the problem of the 
inhomogeneous perturbations,
as well as realistic cosmological scenarios with $F<0$ 
which are explicitly free of singularities \cite{ABG02}.

\vskip 0.7cm

{\noindent \bf Conclusion} 

\vskip 0.3cm

\noindent 
Einstein's gravity with a conformally coupled scalar leads to an 
attractive alternative to usual (minimally coupled) 
inflationary models. New scenarios with a unique dynamics
have been found, which cannot be reduced to minimally 
coupled scalar-driven inflation. However, the stability of these models 
has been a matter of some controversy. Since the conformal factor $F(\psi)$
multiplying the gravitational action can assume negative values 
in these models, it has been suggested that they are unstable 
and therefore must be ruled out \cite{Farese}. 
 
In this paper we have shown that this is not so: although  
the gravitational action indeed acquires a negative sign when $F(\psi)<0$,  
the same happens to the scalar degree of freedom.
As a result, gravitons cannot decay into scalars, or vice versa.

Of course, one should eventually consider some type of matter
besides gravity and the NMC scalar field $\psi$. An additional 
scalar field $\varphi$, for example, would be unstable with respect 
to gravity and the scalar field $\psi$, since the sign of the kinetic 
energy term of $\varphi$ is insensitive to the sign of $F(\psi)$.
The same seems to hold for all bosonic and fermionic fields 
(bosons and fermions do not ``switch'' energy states when the
metric changes lorentzian signature.)
Therefore, the presence of additional (positive-energy) fields at the 
time of inflation, when $F(\psi)<0$, could trigger exactly the types of
instabilities that we tried so hard to avoid\footnote{One way to avoid this
would be to suppose that only gravity and the NMC scalar
field were present at the energy scales corresponding to inflation.}.

Apart from being evidently a potential disaster, this suggests an
interesting, although very speculative, mechanism for reheating the 
Universe after inflation: suppose that initially
all fields were in negative energy states.
This can be achieved by switching the signs of the actions of 
all matter fields except gravity and the NMC scalar.
As inflation nears its end, $F(\psi) \rightarrow 0$ and 
the gravitational and scalar sectors switch 
the signs of their actions due to the effect that we discussed in this
paper. But now this means that a huge instability develops at 
the end of inflation,
between gravity and the inflaton on the one hand, and the rest of
the matter fields on the other hand. The outcome of such an instability
has to be an explosive process which ends in some stable ground state ---
which is presumably the state in which we live until now.

A concrete example can be worked out if we consider gravity, 
the NMC scalar field and fermions.
Suppose that initially the system is in the $F<0$ sector
and all fields are in negative energy eigenstates, so that 
the whole theory is stable. For this
to work out, the fermionic Lagrangian would carry a minus
sign with respect to the usual one.
It can be thought that the fermions initially occupy the 
Dirac sea of positive energy eigenstates, while the negative 
eigenstates are essentially empty (in other words, the Dirac seas 
are reversed with respect to the usual picture.)

As inflation ends and the field passes through the point $\psi^2=1/\xi$, 
the energy eigenstates of the gravitational and scalar degrees of 
freedom become positive, while the fermionic degrees of freedom are
still negative. This instability endures for a very short time,
during which the fermionic states jump out of their positive-energy holes
and fill in the negative eigenstates of the Dirac sea. While this explosive
process happens, much energy can be extracted from the gravitational
and scalar degrees of freedom, and thrown into the fermionic degrees
of freedom. In this manner, the Universe can be filled with radiation --- 
i.e., reheated.

In a forthcoming paper we will show that inflationary 
models which pass from the $F<0$ to the $F>0$ sectors are acceptable 
phenomenologically, provided 
that the transition between these sectors is free of singularities and 
that a final stable ground state can be reached after the transition.
In that paper we will also discuss theories where the whole cosmological
evolution, up to and including the present time, takes place in 
the $F<0$ sector.

\vskip 1cm 
\noindent{\bf Acknowledgments} 
\vskip 0.3cm 

\noindent 
We would like to thank E. Abdalla, S. Deser, G. Esposito-Far\`ese, 
V. Faraoni, V. Mukhanov, D. Polarski, A. Saa, A. Starobinski and
J.-P. Uzan
for valuable discussions. L. R. A. thanks the Physics Department of the  
Universit\'e Libre de Bruxelles for its warm hospitality during 
parts of this project; he also thanks the financial support of 
FAPESP (Brazil). L. B. and E. G. would like to thank the warm hospitality
of USP and UNICAMP (Brazil), where this work was completed.
The authors would like to acknowledge support from 
the EEC Contract \# HPHA-CT-2000-00015
and from the OLAM Fondation Pour La Recherche Fondamentale 
(Brussels).

\end{document}